\documentstyle[aps,prl,preprint]{revtex}
\begin{document}
\draft

\title{Valence band spectroscopy in V-grooved quantum wires}

\author{G. Goldoni, F. Rossi, E. Molinari}
\address{Istituto Nazionale di Fisica della Materia (INFM) and \\
Dipartimento di Fisica, Universit\`a di Modena, Via Campi
213/A, I-41100 Modena, Italy}

\author{A. Fasolino}
\address{Institute of Theoretical Physics, University of Nijmegen,
Toernooiveld, 6525 ED Nijmegen, The Netherlands, and \\
Istituto Nazionale di Fisica della Materia (INFM) and \\
Dipartimento di Fisica, Universit\`a di Modena, Via Campi
213/A, I-41100 Modena, Italy}

\author{R. Rinaldi, R. Cingolani}
\address{Istituto Nazionale di Fisica della Materia (INFM) and \\
Dipartimento di Scienza dei Materiali, Universit\`a di Lecce, Via Arnesano, 
I-73100 Lecce, Italy}

\date{\today}
\maketitle

\begin{abstract}
We present a combined theoretical and experimental study of the 
anisotropy in the
optical absorption of V-shaped quantum wires. By means
of realistic band structure calculations for these structures, we
show that detailed information on the heavy- and light-hole states
can be singled out from the anisotropy spectra
{\em independently of the electron confinement},
thus allowing accurate valence band spectroscopy.

\end{abstract}

\narrowtext

In semiconductor quantum wires, the anisotropic absorption of light 
linearly polarized parallel or perpendicular to the wire axis has long 
been recognized to be the effect of the quasi-one-dimensional (quasi-1D) 
character of electronic states, combined with heavy- and light-hole (HH 
and LH) mixing.\cite{Sercel90,Bockelmann92}
The observation of optical 
anisotropy has since been considered as an evidence of the 1D character 
of nanostructured materials. 
\cite{review} However, 
even in samples showing the best optical properties,
the interpretation of optical spectra is difficult owing to 
the large broadening of the peaks, which is comparable to the typical 
inter-subband splittings, and to
the lack of realistic band structure calculations, so far
limited to wires with model geometries.
\cite{Bockelmann92,Citrin92,Ando93,Goldoni95} 
Moreover, the confinement-induced blueshift is
dominated by the light effective mass of conduction electrons, and 
information concerning hole states are even more difficult to extract 
from the spectra. 

The aim of this letter is to show,
by a combination of optical measurements and 
band structure calculations, that in realistic structures 
the details of the optical anisotropy 
can be traced to HH and LH states and, therefore, allow for a 
{\it spectroscopy of hole states}. 
We shall focus on quasi-1D structures 
obtained by epitaxial growth on non-planar substrates (V-shaped wires)
\cite{Kapon89,Gailhanou93,Tiwari94}, 
which rank among the best available samples from the point of view
of optical properties. 

Our samples are arrays of GaAs quantum wires embedded in 
(GaAs)$_8$(AlAs)$_4$
superlattices (SL), grown by molecular beam epitaxy (MBE) on V-grooved
substrates obtained by holographic lithography and wet etching. 
In our case, the thickness of the crescent-shape wire which self-organizes 
at the bottom of the V-groove ---as obtained by TEM analysis---
is approximately $80\,\mbox{\AA}$. More details on sample parameters 
are reported in Ref.~\cite{Rinaldi94}.

In Fig.~1a we show the low-temperature (10 K) 
photoluminescence excitation (PLE) spectra (thin lines) for a V-shaped wire. 
The spectra are obtained  with 
light polarized along the wire axis and perpendicular to it (and 
to the growth direction). 
We shall focus on the excitonic continuum, i.e., the energy range 
above $1.575\,\mbox{eV}$. The broad peaks in PLE spectra
are very different from the singularities which one would naively expect 
from the ideal joint density of states (DOS)
of quasi-1D electrons and holes, 
because of the relatively small intersubband splittings compared to the
inhomogeneous broadening \cite{Vancouver} and the 
effects of Coulomb interaction \cite{Rossi96}, 
so that the interpretation is not easy. As we shall demonstrate, 
however, additional information
can be gained by considering the optical anisotropy 
between the two polarizations (thick line in Fig.~1a), defined as 
$100*(I_\parallel-I_\perp)/I_\parallel$, where $I_\parallel$, $I_\perp$ are 
the PLE intensities for the two polarizations. 
In this range of energy,  
the anisotropy lies in the range 10-20\%, but goes through a deep 
minimum around 1.59-1.60 eV, where it is almost suppressed. 

We have performed realistic band structure calculations
within the envelope function approximation.
The HH-LH mixing was modelled by the Luttinger Hamiltonian \cite{Luttinger56},
with proper account of crystallographic directions \cite{Xia91}. 
As in Ref.~\cite{Rinaldi94}, the wire potential profile 
was extracted from a TEM micrograph of the sample. 
In the present complex geometry, 
the common approach 
of first solving the Schr\"odinger 
equation for uncoupled HHs and LHs and then using them as a basis set 
to represent the Luttinger Hamiltonian, results in a very large
computation due to the poor convergence with 
respect to the number of subbands included in the basis.
To overcome this difficulty, our approach is to set up a basis set 
from the eigenstates of only one type of particle, with a fictitious mass
to be specified later,
confined in the V-shaped wire potential. These are obtained 
numerically in terms of a plane wave expansion as in 
Ref.~\onlinecite{Rinaldi94}.
In this scheme the additional cost 
in the calculation of matrix elements of the 
Luttinger Hamiltonian is more than compensated, 
because only one type of particle needs to be calculated, and, 
more importantly, the convergence can be improved by 
properly choosing the fictitious mass. 
In practice, we find it efficient to choose a value close 
to the HH mass \cite{fictitious}.

Band parameters are taken from GaAs bulk values. The only unknown 
parameters are the barrier heights for conduction and valence electrons: 
in fact, the embedding short period SL is modeled 
by a homogeneous barrier with an effective band offset. 
The conduction band offset was taken to be $150\,\mbox{meV}$, a value that 
was independently proved to reproduce PLE and magnetoluminescence experiments
\cite{Rinaldi94}. The criterium for the choice of the valence band offset 
is a by-product of the present work and will be discussed below.

The calculated absorption spectra and the relative optical anisotropy, 
obtained from the full band structure 
within the dipole approximation, are shown in Fig. 1b. A gaussian 
broadening of $\pm 4.5\,\mbox{meV}$ has been 
included. The anisotropy 
(thick line) shows a minimum, and then reaches a value of $\sim20\,\%$.
The comparison with experiments is quite favourable for energies  
around the minimum, while the agreement worsens in the high energy range, 
where the calculated anisotropy drops rapidly and 
finally changes sign. This discrepancy might be due to our ``effective'' 
description of the barriers which affects particularly the higher-lying 
hole states. 

As for the calculated absorption spectra,  
apart from a redshift of the continuum onset 
with respect to the experimental curves (which 
is within the uncertainty in the wire geometry),
the main difference lies in the relative intensities of the two broad
features at $1.59\,\mbox{eV}$ and $1.61\,\mbox{eV}$ in the experiment. 
This might be due to the Coulomb interaction, which 
is neglected in the present calculation, and 
causes a change in the excitonic continuum. 
Note, however, that while each single polarization
is significantly changed by excitonic effects, the relative anisotropy 
seems to be much less affected. 

The suppression of the optical anisotropy at $\sim 1.59\,\mbox{meV}$ is 
due to states which are mainly of LH character:
As an indication of the role of LH states, 
we also show in Fig.~1b the calculated absorption spectra for light 
linearly polarized parallel to the growth direction ($z$ direction). 
Although measurements with this polarization are actually difficult
because of the small active volume, it is instructive to calculate,
since $z$-polarized light only couples with 
the LH component of hole states.\cite{Bockelmann92}  
Figure~1b shows that the minimum in anisotropy actually 
lies in the range where the $z$-polarized 
absorption (and therefore the LH character) has a sudden increase.

Due to the large DOS, the electron and
hole states at $k=0$ ($k$ is the 1D wavevector along the wire axis)
are mainly responsible for the features in the 
spectra. Due to the quasi-1D confinement provided by the wire barriers, 
and in contrast to the case of quantum wells, these 
states are of mixed HH and LH character.\cite{Bockelmann92} 
Figure~2 shows the calculated 
HH character of $k=0$ hole states (labeled by the quantum number $n_h$)
vs. the confinement energy.~\cite{foot}
Even in a strongly confined system as the present V-shaped wire, 
the ground state 
is almost a pure HH state (92$\,\%$); the transition 
involving this hole state ($n_h=1$) and the first electron state 
($n_e=1$) is mainly 
responsible for the first peak in the calculated 
spectrum.~\cite{elettroni} 
The excited hole states show an increasingly  
mixed HH-LH character, so that the classification in 
HH and LH states, familiar from quantum wells, becomes inappropriate. 
Simultaneously to the decrease in the HH character, 
the wavefunctions of excited states become more and more 
extended along the ``wings''
of the wire, in analogy to the simpler case of conduction electrons
\cite{Rinaldi94},
and, consequently, they couple to excited electron subbands, mostly 
with $n_e=n_h$);
starting from the $n_h=2$ level in Fig.~2, they contribute to the 
broad features above $\sim1.59$ in the calculated spectra. However, this 
progression is interrupted by the state indicated with an arrow in 
Fig.~2. This state is of mainly LH character (56\,\%), and 
for brevity it will be indicated below as {\em the} LH state. Although 
this is a highly excited state ($n_h=13$), its
wavefunction, shown in the insets, is localized 
in the center of the wire. Therefore, due to the large overlap,
this state couples primarily with 
the first conduction subband ($n_e=1$), 
making a contribution to the low energy 
part of the spectrum. The large LH component, moreover, makes that the 
intensity for the two polarizations is reversed with respect 
to the strongly HH ground state, causing the dip in the anisotropy 
at $1.59\,\mbox{meV}$.

An immediate consequence of the above reasoning is that,
since both the ground HH state and the LH state couple 
with the lowest electron subband, 
the difference in energy between the onset of the continuum 
and the position of the dip in the anisotropy is a direct measure of 
the energy splitting between the ground HH and the first LH states, 
{\em independently of the  electron confinement}. 
This splitting, which in the experiment of Fig.~1a is $\sim 16\,\mbox{meV}$, 
has been used in our calculation to estimate
the effective valence band offset; its value was finally taken as
$85\,\mbox{meV}$.\cite{offset} 

In conclusion we have shown by realistic 
band structure calculations that the optical  anisotropy 
of V-shaped quantum wires contains detailed information on the valence 
subbands, which can be singled out from the conduction electron 
contribution,  allowing for a spectroscopy of the hole states.

\begin{figure}
\caption{a) 
Low-temperature 
PLE spectra measured with incoming light linearly polarized
perpendicular to the wire axis (thin solid line), parallel to the wire
axis (dotted line), 
and relative anisotropy (thick solid line). 
b) Calculated absorption spectra with light linearly polarized
perpendicular to the wire axis (thin solid line), parallel to the wire
axis (dotted line), parallel to the $z$ direction (dashed line), and
relative anisotropy (thick solid line).} 
\end{figure}

\begin{figure}
\caption{Percentage of HH character for the lowest hole subbands at
$k=0$ vs subband energy. 
The arrow indicates the first state which is mainly ($56\,\mbox{\%}$) of LH
character. Insets: total charge density, and HH- and
LH-projected charge densities for the state indicated with the arrow 
($n_h=13$).
Darker regions
correspond to larger values of the charge density; thin lines represent 
the potential profile used in the calculation. In-plane axis  
are in nm (note the different scale in the two directions). For clarity, 
in the three panels grey levels are set to different scales.} 
\end{figure}



\widetext

\begin{references}
\bibitem{fictitious} The choice of such fictitious mass is purely 
dictated by computational convenience: once convergence is achieved, 
it does not affect the final results in any way.
\bibitem{foot} 
The angular momentum quantization axis is chosen along the direction of 
strongest confinement, i.e. the [001] growth direction.
\bibitem{elettroni} To help with
the interpretation of the theoretical spectra 
we record here the electron confinement energy of the three lowest 
subbands $n_e=1,2,3$: 
$43.3\,\mbox{meV}$, $57.3\,\mbox{meV}$, and $65.3\,\mbox{meV}$. 
The band gap is $1.519\,\mbox{eV}$.
\end{references}
\end{document}